\documentstyle[epsf,12pt,psfig]{article}
\begin{document}

\begin{center}

{\large\bf THE BOTTOM BARYON PAIR PRODUCTION
IN THE $e^+e^-$-ANNIHILATION}
\vspace{3mm}

{\it V.A. Saleev\footnote{E-mail address:
                             saleev@info.ssu.samara.ru}\\
%
Samara State University, Samara 443011, Russia}

\end{center}


\begin{abstract}
In the framework of the nonrelativistic diquark model of
the heavy baryons and the perturbative approach of QCD we
predict the value and energy dependence for the total cross
section of the $\Lambda_b$- and $\Sigma_b^*$-baryons
pair production at the energy range of the KEKB and PEP-II
$e^+e^-$-colliders.
\end{abstract}

\section{Introduction}

At the present time the existence of the heavy baryons containing
b-quark is confirmed experimentally with the high order of accuracy.
The mass, the life time and the main partial widths of the decays are
measured for the lightest of the bottom baryon $\Lambda_b$
($I(J^P)=0(\frac{1}{2}^+)$) in the experiments at $p\bar p-$colliders
CERN \cite{1} and FNAL \cite{2} as well as at $e^+e^--$collider LEP
\cite{3} in the Z-boson pole. Recently the first evidence of the
$\Sigma_b,\Sigma_b^*$ \cite{4} and $\Xi_b$ \cite{5} baryons was obtained at
LEP too. The great surprise is to be a large depolarization of the
$\Lambda_b$ baryons produced in the Z boson decays
${\cal P}_{\Lambda_b}=-0.23^{+0.24}_{-0.20}$ \cite{6},
which contradict the naive predictions \cite{7} that the main part of
the initial polarization of the bottom quark to transfer to the $\Lambda_b$
baryon ${\cal P}_{\Lambda_b}\approx {\cal P}_b=-0.94$. Howere, the accuracy
of the obtained results on the bottom baryon production at the high
energy $p\bar p$ and $e^+e^-$ colliders isn't enough for the precise
testing the theoretical predictions.

The main reasons of it are the small luminosity of the high energy
colliders ${\cal L}\sim 10^{30}$ sm$^{-2}$s$^{-1}$ and the nonresonant
nature of the heavy quark production processes. By means of this
point, the more careful conditions for the study of the b-quark
baryons are waiting at the future B-factories KEKB and PEP-II at the
energy range of $e^+e^--$collisions is about of 11-14 GeV and the high
luminosities  ${\cal L}\sim 10^{34}$sm$^{-2}$s$^{-1}$.
The exclusive bottom baryon-antibaryon pair production is a very
interesting subject for the study of the heavy baryon physics at the
B-factories. The cross sections, as a function of the total $e^+e^-$
beam energy $\sqrt{s}$, of these processes have a resonance like
nature and clean experimental signatures. With the theoretical point
of view the processes of the exclusive baryon pair creation near the
threshold of the bottom quarks production are more calculable in the
framework of QCD and the quark-diquark model of the baryons.

Here we study the processes of the bottom baryon-antibaryon spin-3/2
and spin-1/2 pair production in the $e^+e^--$annihilation at the
energies of KEKB and PEP-II $e^+e^--$colliders:
$e^+e^-\to\Lambda_b\bar\Lambda_b$ and $e^+e^-\to \Sigma_b^*\bar\Sigma_b^*$.

\section{The model}

Recently \cite{8,9} we have suggested the approach based on the
perturbative QCD and the nonrelativistic diquark model of the baryons,
which let us to calculate the cross sections and spin asymmetries
for the heavy and doubly heavy baryon production. In the frame work of
the diquark model of the baryons \cite{10} the heavy baryon may be
presented as a system of the heavy quark Q and the diquark D.
In the case of the bottom baryon it has
$m_Q,m_D>>\Lambda_{QCD}$ ($m_Q\sim 5$ GeV, $m_D\sim 1$ GeV)
and we can use the perturbative expansion in
$\alpha_s$, taking into account the nonperturbative effects by diquark
form factors. Because of the reduced mass of the heavy quark - diquark
system is large in compare with the binding energy (it is the
same order as for charmonium),
we shell use the nonrelativistic approximation for the description of
the quark-diquark transition into the baryon.
In the approximation of the zero binding energy and small relative
velocity of the constituents, the probability of the quark-diquark
transition into the baryon is determined by the baryon wave function in
the origin $\Psi(0)$ \cite{10b}. This parameter may be calculated using the
potential approach with the quark-diquark potential which have been
fixed in the calculating of the heavy baryon masses \cite{11}.

The amplitudes of the processes $e^+e^-\to\Lambda_b\bar\Lambda_b,\Sigma_b^*\bar\Sigma_b^*$ are
described by Feynman diagrams which are shown in Fig.1. The
contribution of the diagrams 3 and 4 may be omitted because of the
strong suppressions due to the diquark form factor in the
photon-diquark vertexes \cite{12}.

In such a way, the virtual photon create the bottom
quark-antiquark pair (diagrams 1 and 2 in Fig.1), which catch the
scalar or the vector diquark, correspondingly.
The gluon couplings to scalar and vector diquarks are presented by the
following expressions \cite{10}:
\begin{equation}
 S_{\mu}^a=-ig_sT^a(p_D'-p_D)_{\mu}F_s(k^2)
\end{equation}

$$V_{\mu}^a=ig_sT^a \Bigl [\varepsilon^*_D\varepsilon_D' (p'_D-p_D)_{\mu}F_1(k^2)+
 \bigl [(p_D\varepsilon_D' ){\varepsilon^*_D}_{\mu}-(p'_D\varepsilon^*_D){\varepsilon_D'}_{\mu}
\bigr ]F_2(k^2)$$
\begin{equation}
+(\varepsilon_D' p_D)(\varepsilon^*_D p'_D)
(p_D-p'_D)_{\mu}F_3(k^2)\Bigr ],
\end{equation}
where $T^a$ are Gell-Mann matrices, $\varepsilon^*_D$ and $\varepsilon_D'$ are the diquark
polarization vectors, $p_D=rp$ and $p_D'=rp'$ are diquark four-momenta,
$F_s,F_1,F_2$ and $F_3$ are form factors depending on the momentum
transfer squared
$k^2=(p_D+p_D')^2=r^2s$.
At first approximation the form factors don't depend on the baryon
type and may be parameterized as the same \cite{12}, where the authors
fit successfully the angular distributions of the baryons in the
processes $\gamma\gamma\to p\bar
p,\Lambda\bar\Lambda$ and $J/\Psi\to p\bar p,\Lambda\bar \Lambda$.
We shall use the following set of form factors:
\begin{eqnarray}
&&F_s(k^2)=\frac{Q_s^2}{Q_s^2-k^2},\\
&&F_1(k^2)=\left (\frac{Q_v^2}{Q_v^2-k^2}\right )^2,\\
&&F_2(k^2)=(1+\kappa)F_1(k^2),\\
&&F_3(k^2)\approx 0,
\end{eqnarray}
where $Q_s^2=3.22$ GeV$^2$, $Q_v^2=1.50$ GeV$^2$ and $\kappa=1.39$
being the anomalous chromomagnetic moment of the vector diquark,
and all form factors are restricted to value smaller than 1.3.

In the case of the $\Lambda_b$-baryon production it has fusion of the heavy
quark and scalar diquark. After some simplifications we have obtained:
\begin{equation}
{\cal M}(e^+e^-\to \Lambda_b\bar\Lambda_b)=\frac{4}{9}
\left [\frac{\Psi(0)}{2m_D}\right ]^2
\frac{(ee_b)^2g_s^2}{r^2s^3}L_{\alpha}H_{1/2}^{\alpha},
\end{equation}
where
$$L_{\alpha}=\bar V(p^+)\gamma_{\alpha}U(p^-),$$
$$H_{1/2}^{\alpha}=\bar U(p)\biggl [(\hat p-\hat p')(\hat p+\hat p_D'+m_Q)
\gamma^{\alpha}+\gamma^{\alpha}(m_Q-\hat p'-\hat p_D)(\hat p-\hat p')
\biggr ]V(p'),$$
$e_b=1/3$, $e^2=4\pi\alpha$, $g_s^2=4\pi\alpha_s (m_D)$, $4/9$ is the
colour factor of the amplitude. The spinors $\bar U(p)$ and $V(p')$
describe $\Lambda_b-$baryons in the final state.

The amplitude for the $e^+e^-$-annihilation into spin-3/2
$\Sigma_b^*$-baryons, corresponding to fusion of the heavy quark and the
vector diquark, can be written as follows:
\begin{equation}
{\cal M}(e^+e^-\to \Sigma_b^*\bar\Sigma_b^*)=\frac{4}{9}
\left [\frac{\Psi(0)}{2m_D}\right ]^2
\frac{(ee_b)^2g_s^2}{r^2s^3}L_{\alpha}H_{3/2}^{\alpha},
\end{equation}
where
$$H_{3/2}^{\alpha}=\bar\Psi_{\sigma}(p)
\biggl [\gamma^{\mu}(\hat p+\hat p_D'+m_Q)
\gamma^{\alpha}+\gamma^{\alpha}(m_Q-\hat p'-\hat p_D)\gamma^{\mu}
\biggr ]\Psi_{\lambda}(p')\cdot$$
$$\biggl\{F_1(k^2)(p'-p)_{\mu}g_{\sigma\lambda}+F_2(k^2)
\bigl [p_{\lambda}g_{\mu\sigma}-p'_{\sigma}g_{\mu\lambda}\bigr ]
\biggr\},$$
$\bar\Psi_{\sigma}(p)$ and $\Psi_{\lambda}(p')$ are the
Rarita-Schwinger spinors for the spin-3/2 baryons.

\section{The results}
The differential cross section for the processes
 $e^+e^-\to\Lambda_b\bar\Lambda_b,\Sigma_b^*\bar\Sigma_b^*$  can be written as follows:
\begin{equation}
\frac{d\sigma}{dt}=\frac{\overline{|{\cal M}|^2}}{16\pi s^2}.
\end{equation}
The total cross section  $\sigma(s)$ will be obtained after
integration over $t$ in limits:
$$t_{min}=M^2-\frac{s}{2}(1-v),\quad t_{max}=M^2-\frac{s}{2}(1+v).$$
where $v=\sqrt{1-4M^2/s}.$ The integrating can be made analytically,
we have found that
\begin{equation}
\sigma(e^+e^-\to\Lambda_b\bar\Lambda_b)=\frac{4\pi}{2187}e_b^2\alpha^2\alpha_s^2
\frac{|R(0)|^4}{m_D^6}\frac{v}{s}\Phi_s(r,v)F_s^2(k^2),
\end{equation}
\begin{eqnarray}
&&\sigma(e^+e^-\to\Sigma_b^*\bar\Sigma_b^*)=\frac{4\pi}{2187}e_b^2\alpha^2\alpha_s^2
\frac{|R(0)|^4}{m_D^6}\frac{v}{s}\cdot\nonumber\\
&&\biggl [\Phi_{11}(r,v)F_1^2(k^2)
+\Phi_{22}(r,v)F_{22}^2(k^2)+\Phi_{12}(r,v)F_1(k^2)F_2(k^2)\biggr ],
\end{eqnarray}
where $R(0)=\Psi(0)/\sqrt{4\pi}$ is the radial part of the baryon wave
function in the origin,
\begin{eqnarray}
&&\Phi_s(r,v)=\frac{9}{16}\biggr [ v^8r(r-2)+2v^6(-3r^2+6r-4)+\nonumber\\
&&4v^4(3r^2-6r+7)+2v^2(-5r^2+10r-16)+3(r^2-2r+4)\biggl ]\\
&&\Phi_{11}(r,v)=\frac{1}{4}\biggr
[3v^8r(r-2)+2v^6(-11r^2+22r-12)+\nonumber\\
&&4v^4(15r^2-30r+29)+2v^2(-37r^2+74r-104)+33(r^2-2r+4)\biggl ]\\
&&\Phi_{22}(r,v)=2\biggr [v^8r(r-1)+4v^6(-2r^2+2r-1)+\nonumber\\
&&v^4(23r^2-23r+22)+4v^2(-7r^2+7r-10)+12(r^2-r+2)\biggl ]\\
&&\Phi_{12}(r,v)=v^8r(r-2)+v^6(_9r^2+18r-8)+\nonumber\\
&&v^4(29r^2-58r+52)+3v^2(-13r^2+26r-36)+18(r^2-2r+4).
\end{eqnarray}

The results of our calculations are presented in the Table 1 for the
maximum values of the total cross sections and ratios:
\begin{equation}
R_{\Lambda,\Sigma}=\frac{\sigma (e^+e^-\to\Lambda_b\bar\Lambda_b,\Sigma_b^*\bar\Sigma_b^*)}{\sigma
(e^+e^-\to b\bar b)},
\end{equation}
where
\begin{equation}
\sigma(e^+e^-\to b\bar b)=3e_b^2\frac{4\pi\alpha_s^2}{3s}\biggl (
1+\frac{2m_Q^2}{s}\biggr )\sqrt{1-\frac{4m_Q^2}{s}}.
\end{equation}

We have found that the value of the cross section for
$e^+e^--$annihilation into heavy baryons strongly depends on the
diquark mass. Roughly speaking
$\sigma(e^+e^-\to \Lambda_b\bar\Lambda_b,\Sigma_b^*\bar\Sigma_b^*)\sim 1/m_D^6$.
The value of the constituent diquark mass $m_D$ is known only
approximately. We shall use $m_D=0.6$ GeV \cite{12} and $m_D=0.9$ GeV.
The second one is more realistic in the case of bottom baryons. We
want note that the mass of the diquark may be fixed by the peak
position in the cross section of heavy baryon pair production in
$e^+e^--$annihilation. So, we have found that at $m_D=0.9$ GeV the
peak position is at $\sqrt{s}=12$ GeV, but at $m_D=0.6$ GeV the peak
position is at $\sqrt{s}=15.5$ GeV. The measurement of the ratio
$R_{\Lambda,\Sigma}$ as a function of $\sqrt{s}$ is the direct way to
study of the asymptotic behaviour of the diquark form factors.
Fig.2 shows the result of the calculation for the
$R_{\Lambda}(\sqrt{s})$ at $m_D=0.9$ GeV. The curver 1 corresponds to
the form factor $F_s(k^2)$ by formula (3), the curver 2 corresponds to
$F_s(k^2)=max\{F_s(k^2)\}=1.3$. Because of the gluon virtuality
$k^2\ge 4m_D^2$ is the same order as $Q_s^2$, the ratio $R_{\Lambda}$
strongly depends on the value of $Q_s^2$ as well as on the shape of
the diquark form factor $F_s(k^2)$. Fig.2 demonstrates that the
asymptotic behaviour of $R_{\Lambda}$ may be fixed at $\sqrt{s}>15$
GeV.

We have found (Table 1), that the cross section for $\Sigma_b^*$-baryon
pair production is larger than cross section for $\Lambda_b-$baryon pair
production:\\
$\sigma(\Sigma_b^*\bar\Sigma_b^*)/\sigma(\Lambda_b\bar\Lambda_b)\approx 30$ at $m_D=0.6$ GeV
and
$\sigma(\Sigma_b^*\bar\Sigma_b^*)/\sigma(\Lambda_b\bar\Lambda_b)\approx 10$ at $m_D=0.9$ GeV.
Our results are in a qualitative agreement with the conclusions of
Ref. \cite{13}, where the
single charmed baryon production at CLEO energies have been calculated
a similar way.

Taking into account that high spin heavy baryons decays into the
ground state $\Sigma_b^*\to\Lambda_b+\pi$, we predict the large difference between
direct and cascade $\Lambda_b\bar\Lambda_b$ pair production in  $e^+e^--$annihilation,
which may be measured experimentally.
We have estimated that the total cross section of the bottom baryon
pair production in the $e^+e^--$annihilation:
$\sigma(\Lambda_b\bar\Lambda_b)+ \sigma(\Sigma_b\bar\Sigma_b)+
\sigma(\Sigma_b^*\bar\Sigma_b^*)=1-16$ pb at the different choice of the diquark mass.
We put here that $\sigma(\Sigma_b\bar\Sigma_b)=\sigma(\Sigma_b^*\bar\Sigma_b^*)/2$.

\section*{Acknowledgments}
We are grateful to  S.~Gerasimov, V.~Galkin, A.~Likhoded and R.~Faustov
for discussions the problems of the heavy
baryon physics.

\newpage

\newpage
\section*{Table 1}
\vspace{5mm}
\begin{center}

\begin{tabular}{|c|c|c|c|c|} \hline
Baryons & $m_D$, GeV & $|R(0)|^2$, GeV$^3$ &$\sigma_{max}$, pb&
$R_{max}$, \% \\ \hline
$\Lambda_b\bar\Lambda_b$ & 0.6 & 0.73 & 0.36 & 0.23 \\ \hline
$\Sigma_b^*\bar\Sigma_b^*$ & 0.6 & 0.73 & 11.0 & 6.4  \\ \hline
$\Lambda_b\bar\Lambda_b$ & 0.9 & 1.2 & 0.05 & 0.03 \\  \hline
$\Sigma_b^*\bar\Sigma_b^*$ & 0.9 & 1.2 & 0.64 & 0.37 \\ \hline
\end{tabular}
\end{center}

\vspace{2cm}
\section*{Figure captions}

\begin{enumerate}
\item  Diagrams used for description of the processes
$e^+e^-\to \Lambda_b\bar\Lambda_b,\Sigma_b^*\bar\Sigma_b^*$.
\item The ratio $R_{\Lambda}$ as a function of $\sqrt{s}$.
The curver 1 corresponds to form factor $F_s(k^2)$ from [13],
curver 2 corresponds to $F_s(k^2)=max\{F_s(k^2)\}=1.3$.
\end{enumerate}

\begin{figure}[p]

\[\psfig{figure=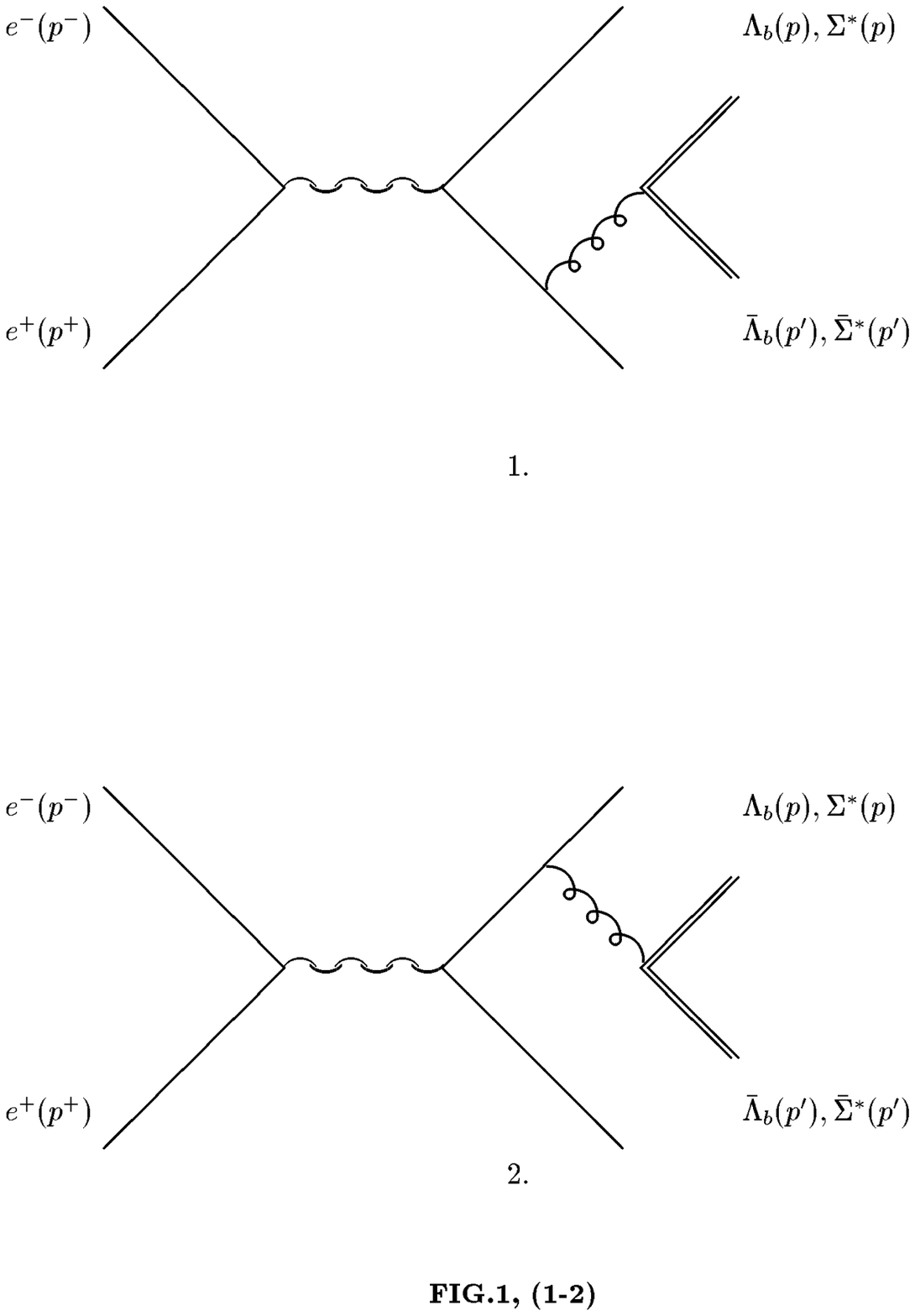,%
     bbllx=2.5cm,bblly=18cm,bburx=18cm,bbury=25cm,%
        width=8cm}
\]
\vspace*{5cm}
\end{figure}

\begin{figure}[p]

\[\psfig{figure=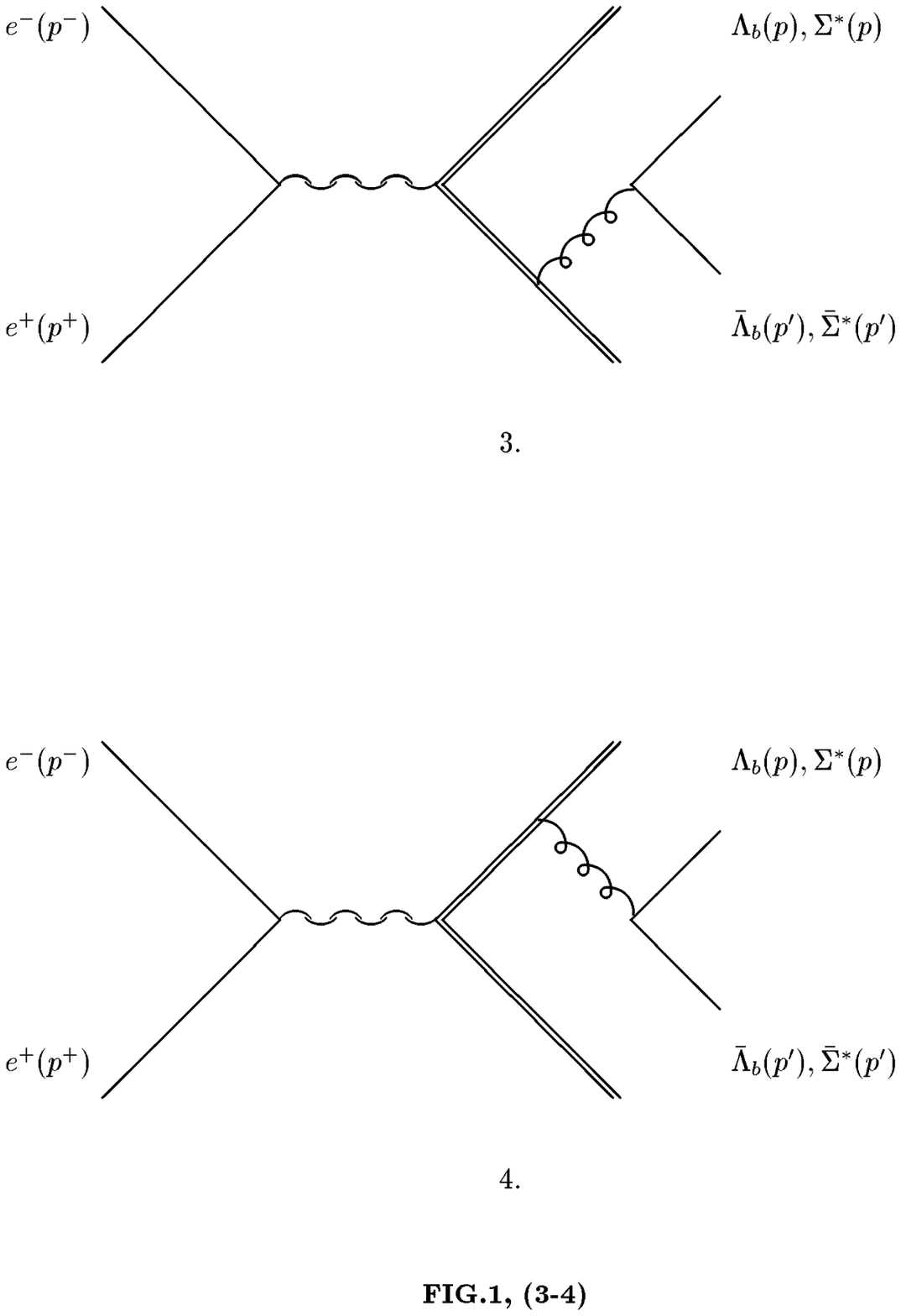,%
     bbllx=2.5cm,bblly=18cm,bburx=18cm,bbury=25cm,%
        width=8cm}
\]
\vspace*{10cm}
\end{figure}

\begin{figure}[p]

\[\psfig{figure=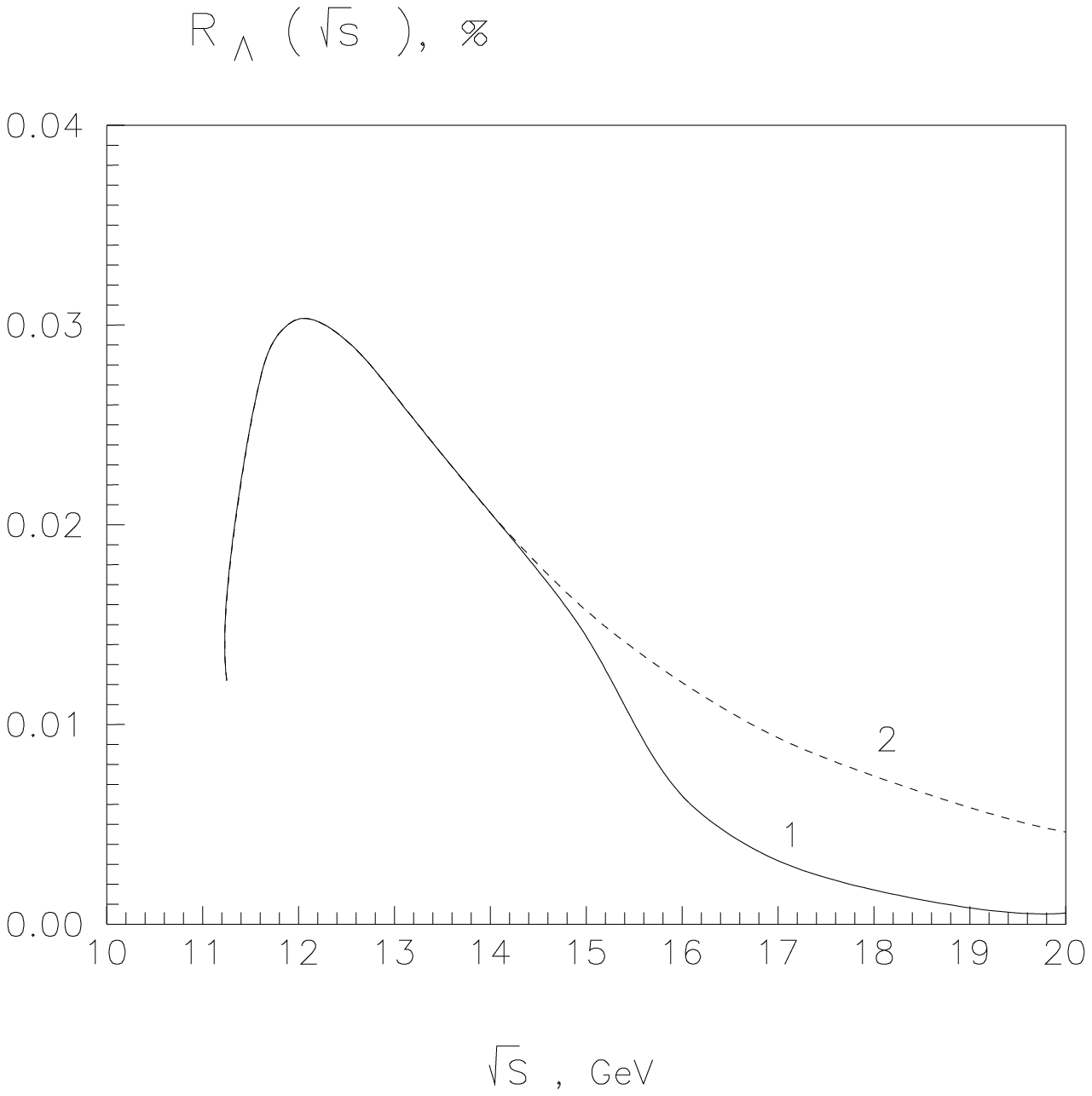,%
     bbllx=2.5cm,bblly=18cm,bburx=18cm,bbury=25cm,%
        width=8cm}
\]

\end{figure}

\end{document}